\documentclass[aps,prc,showpacs,preprintnumbers,twocolumn,amsmath,nofootinbib,amssymb,floatfix]{revtex4}
\usepackage{graphicx,dcolumn,booktabs,bm}
\usepackage{longtable}
\usepackage{threeparttable}
\usepackage{txfonts}
\usepackage{overpic}
\usepackage{float}
\usepackage{amssymb}
\usepackage{epstopdf}
\usepackage{indentfirst}
\usepackage{feynmf}    %{feynmp}
\usepackage{slashed}  %for Feynman symbolsf
\usepackage{cases}
\usepackage{ulem}
\usepackage{color}
\usepackage{array}
\usepackage{booktabs}
\usepackage{multirow}
\usepackage{epsfig,dsfont,amsmath,amsfonts,amsbsy,mathrsfs}
\usepackage{cancel}
\usepackage{xcolor}  

\usepackage[colorlinks, citecolor=blue,anchorcolor=red,menucolor=red, linkcolor=red,filecolor=red,runcolor=red,urlcolor=blue,frenchlinks=red]{hyperref}
\makeatletter
\@addtoreset{equation}{section}
\makeatother

\allowdisplaybreaks
\begin{document}
\title{WKB approximation under Killingbeck potential and its application to double-heavy meson spectroscopy}
\author{{Cheng-Qun Pang}$^{1}$\footnote{Corresponding author}}\email{xuehua45@163.com}
\author{{Lhamo Chosto}$^{2,3}$}\email{lamaoyy@163.com}
\author{{Ya-Rong Wang}$^{4}$}\email{nanoshine@foxmail.com}
\author{{Zhi-Bin Gao}$^{5}$}\email{GZB2807256729@163.com}
\author{{Hao Chen}$^{3,6}$}\email{chenhao\_qhnu@outlook.com}
\author{{Yu-Long Kang}$^{3,6}$}\email{kungyulong@163.com}
\author{{Wei Li}$^{3}$}\email{liweibb@126.com}
\affiliation{
$^1$School of Physics and Optoelectronic Engineering, Ludong University, Yantai 264000, China\\
$^2$Gonghe County Senior High School, Qinghai Province, Gonghe 811800, China\\
$^3$College of Physics and Electronic Information Engineering, Qinghai Normal University, Xining 810000, China\\
$^{4}$Center for Joint Quantum Studies and Department of Physics, School of Science, Tianjin University, Tianjin 300350, China\\
$^5$School of General Education, Qinghai Polytechnic University, Xining 810000, China\\
$^6$Academy of Plateau Science and Sustainability, Xining 810016, China\\}
\begin{abstract}
    It is well known that non-relativistic or semi-relativistic constituent quark models can describe  a large number of the meson and baryon properties with surprising accuracy. In this work, we study  Killingbeck potential using  th  WKB approximation in the constituent quark model. The implicit approximate  solutions for   the energy  eigenvalues, together with analytical expressions for the root-mean-square radii and the probability densities at the origin,  of the Schr\"{o}dinger equation   are  obtained by using appropriate quantization conditions. Finally, we discuss  applications to double-heavy meson spectroscopy within the Killingbeck potential. 
\end{abstract}

\maketitle
\section{introduction}\label{s1}

In the past decades, many potential energy functions that play important roles in describing physical phenomena have been proposed.  Some wave equations with special potentials are exactly solvable, whereas many physically relevant potentials require numerical or approximate analytical methods~\cite{1975Dynamics,Brau_2000,Sergeenko_1996}. As a semiclassical approximation method, the Wentzel-Kramers-Brillouin (WKB) approximation was originally developed by Wentzel, Kramers, and Brillouin in 1926~\cite{Wentzel1926,Kramers1926,Brillouin1926}. Later, it was successfully applied to particle energy spectra in high energy physics by J. S. Kang~\cite{1975Dynamics,1975Is} and C. Quigg \textit{et al.}~\cite{Quigg:1977wn}. This approximation can be used to analyze the Schr\"{o}dinger equation and has been successfully applied to many important problems~\cite{Sergeenko_1996,Chebotarev1998Extensions,Robinett2009The,Al-Masaeed:2021ske,2002A}. The WKB approximation based on the Schr\"{o}dinger equation is expected to give reliable energy eigenvalues in the large-quantum-number limit~\cite{1995Introduction}. For low-lying states the accuracy is model dependent, and the WKB results should be checked against exact or high-precision numerical solutions.  

In 2005, Z. Q. Ma \textit{et al.} proposed an accurate quantization condition for one-dimensional quantum system and three-dimensional spherically symmetric quantum system~\cite{Ma2005,MA_2005}. F. Brau reported the relationship among energy, radial quantum number, and orbital angular momentum~\cite{Brau_2000}. H. F. Lalus \textit{et al.} used the WKB method to analyze the physical conditions that the system must satisfy in the process of tunneling~\cite{2021WKBa}.
E. Omugbe \textit{et al.}, obtained the non-relativistic ro-vibrational energy spectra, expectation values, and the thermodynamic properties of the Deng Fan potential within the framework of the WKB approximation method~\cite{Omugbe_2020}. Moreover, the authors in Refs.~\cite{Energy_2020,2006Analytical,
Omugbe:2020oaq,2012Bohr,2021Approximate,
2004Simulation,2019Calculation} have used WKB method to study mass spectra under various potential energies.
The bound state solution to the wave equations under the quark-antiquark interaction potential functions such as the Killingbeck or the Cornell potential have attracted much research interest in high energy physics~\cite{Omugbe:2020oaq}. In Refs.~\cite{Brau_2000,Olsson:1996jg,2000From}, the authors used WKB method to study the trajectory relation with the assumption of linear confinement potential. Compared with the Cornell potential, the Killingbeck potential contains an additional quadratic term and has been used to study double-heavy meson spectra~\cite{Kumar:2013dsa,Abu-Shady:2015aka}. The advantage of the present WKB treatment lies in its analytical nature: it allows the energy spectrum and wave-function observables, such as the root-mean-square (RMS) radius and the probability density at the origin, to be followed for arbitrary radial and orbital excitations. It also provides a controlled way to examine how a quadratic correction affect heavy-quark potentials. However, direct WKB studies of the mass spectra and wave-function observables generated by the Killingbeck potential remain limited. This work is intended to address that problem.

Reference \cite{ovsiyuk2012schrodinger} presents several exactly solvable forms of the Killingbeck potential. These exact analytical results provide valuable benchmarks for assessing the accuracy of approximate methods, such as WKB method. This motivates the use of complete elliptic integrals to obtain analytical expressions for physical observables and to track hadron properties without repeatedly solving the differential equation numerically. The main focus of this paper is to obtain WKB solutions for the Killingbeck potential in the constituent quark model. Furthermore, the RMS radii and the probability densities at the origin, which are closely related to short-distance decay properties, are obtained and applied to double-heavy mesons. 

This paper is organized as follows. In Sec.~\ref{s2}, the WKB eigenvalues, RMS radii, and probability densities at the origin of the Killingbeck potential are introduced. In Sec.~\ref{s3}, we compare the WKB approximate solutions with quasi-exactly solvable benchmarks for the Killingbeck potential. In Sec.~\ref{s4}, we analyze the mass spectra and spatial observables of double-heavy mesons and compare the WKB results with other theoretical calculations and available experimental data. A discussion and conclusion are given in Sec.~\ref{s5}.

\section{WKB approximation}\label{s2}
F. Brau calculated the energy spectrum under Cornell potential in the quark model and reported the relation among energy, radial quantum number, and orbital angular momentum~\cite{Brau_2000}. Following this strategy, we solve the spectra and spatial observables of double-heavy mesons under Killingbeck potential. 

The basic quantities in the Bohr-Sommerfeld quantization are the action variables~\cite{Brau_2000}:
\begin{eqnarray}
  J_s=\oint{p_s\textrm{d}q_s}, \label{1}
\end{eqnarray}
where $s$ labels the degrees of freedom of the system, and $q_s$ and $p_s$ are the generalized coordinate and the corresponding conjugate momentum, respectively. The integral is performed over one cycle of the motion. The action variables are quantized according to the prescription
\begin{eqnarray}
J_s=(n_s+c_s)h, \label{22}
\end{eqnarray}
where $h$ is Planck's constant, $n_s\geq0$ is an integer quantum number and $c_s$ is a real constant that should be set to $\frac{1}{2}$ according to Langer \cite{Brau_2000}.

Under the WKB approximation, the non-relativistic Hamiltonian corresponding to the Killingbeck potential reads, in natural units ($\hbar=\frac{h}{2\pi}=c=1$), as:
\begin{eqnarray}
\begin{split}
H&= \frac{1}{2\mu}(p^2_{\rho} + \frac{p^2_{\phi}}{\rho^2})- \frac{\kappa}{\rho} + \tau \rho + \nu \rho^2
\\&+ m_1 + m_2 + C + \frac{\sigma }{m_1 m_2}\bm S_1\cdot\bm S_2,\label{11}
\end{split}
\end{eqnarray}
{with the spin-spin expectation value being}
\begin{eqnarray}
\langle \bm{S_1}\cdot\bm{S_2}\rangle=\frac{1}{2}S(S+1)-\frac{3}{4}.\label{4}
\end{eqnarray}
Here, $\mu$ is the reduced mass; {$\kappa$, $\tau$, $\nu$, $C$, and $\sigma$} are the parameters to be fitted; $m_1$ and $m_2$ are the masses of the two quarks. In addition,  $p_\phi$ can be identified with $L$ ($p_\phi=L$), where $L=l+c_\phi$ and $l$ is the orbital angular momentum quantum number. The radial momentum $p_{\rho}$ is derived from the conservation of the shifted energy $\lambda^\prime$ ($\lambda^\prime=\lambda-m_1-m_2-C
-\frac{\sigma}{m_1m_2}\left[\frac{1}{2}S(S+1)-\frac{3}{4}\right]$, where $\lambda$ is the total energy of the system):
\begin{eqnarray}
p_{\rho}= \frac{1}{\rho}\sqrt{-2\mu \nu \rho^4-2\mu \tau \rho^3+2\mu \lambda^\prime \rho^2+2\mu \kappa \rho-L^2 }.\label{5}
\end{eqnarray}

With the replacements
\begin{equation}
\label{schrt}
\rho\to\frac{r}{2\mu \kappa}, \tau\to4\mu^2 \kappa^3 a, \nu\to8\mu^3 \kappa^4b,  
\lambda^\prime\to2\mu\kappa^2 E,
\end{equation}
The Hamiltonian given in Eq. (\ref{11}) can be transformed into the dimensionless form
\begin{equation}
    H^\prime = p^2_{r} + \frac{p^2_{\phi}}{r^2}- \frac{1}{r} + ar + br^2.
    \label{15}
\end{equation}

\subsection{WKB approximation eigenvalue of  Killingbeck potential.}
The radial momentum $p_r$ derived from the effective Hamiltonian $H^\prime$ is given by
\begin{equation}
p_r(r) = \pm\sqrt{ E - \left( \frac{L^2}{r^2} - \frac{1}{r} + ar + br^2 \right) }.
\end{equation}
By factoring out $1/r$, the momentum can be  expressed via a quartic polynomial $P_4(r)$:
\begin{equation}
\label{p4}
p_r(r) = \frac{\sqrt{P_4(r)}}{r}, \quad P_4(r) = -b r^4 - a r^3 + E r^2 + r - L^2.
\end{equation}

The four analytical roots $R_1$, $R_2$, $R_3$, and $R_4$ of the momentum polynomial $P_4(r) = 0$ can be determined using algebraic invariants. We define the fundamental resolvent parameters $D_2$, $D_1$, $\Delta_0$, and $\Delta_1$ as follows:
\begin{align}
D_2 &= \frac{-8bE - 3a^2}{8b^2}, \\
D_1 &= \frac{a^3 + 4abE - 8b^2}{8b^3}, \\
\Delta_0 &= E^2 + 3a + 12bL^2, \\
\Delta_1 &= 2E^3 + 9aE - 27a^2L^2 - 27b - 72b E L^2,
\end{align}
$Q$ and $Z$ can be given as:
\begin{align}
Q &= \sqrt[3]{\frac{\Delta_1 + \sqrt{\Delta_1^2 - 4\Delta_0^3}}{2}}, \\
Z &= \frac{1}{2}\sqrt{-\frac{2}{3}D_2 - \frac{1}{3b}\left(Q + \frac{\Delta_0}{Q}\right)}.
\end{align}
Then, the exact analytical roots  are explicitly formulated as:

\begin{align}
 R_{1,2,3,4}= -\frac{a}{4b} + \varepsilon_1 Z + \frac{\varepsilon_2}{2}\sqrt{-4Z^2 - 2D_2 - \varepsilon_1\frac{D_1}{Z}}, 
\end{align}
with $
 \varepsilon_1 = \pm 1,\ \varepsilon_2= \pm 1$.

If $P_4(r)$ possesses four real roots ordered as $R_1 > R_2 > R_3 > R_4$, the two non-negative roots define the classical turning points, $r_+ = R_1$ and $r_- = R_2$.

The Bohr-Sommerfeld quantization condition, $\int_{r_-}^{r_+} p_r \textrm{d}r = (n + c_r)\pi$, can be expressed as:
\begin{equation}
-b Z_3 - a Z_2 + E Z_1 + Z_0 - L^2 Z_{-1} = \left(n_r +c_r\right)\pi,
\label{eq:BS_condition}
\end{equation}
where $Z_k = \int_{r_-}^{r_+} r^k P_4^{-1/2}(r) \, \textrm{d}r$ and $n_r$ is the radial quantum number.

In order to  express $Z_k$ as special functions,
we define the purely algebraic geometric parameters:
\begin{align}
\delta&= (R_2-R_3), \\
g &= \frac{2}{\sqrt{b (R_1-R_3)(R_2-R_4)}}, \\
\eta &= \frac{(R_1-R_2)(R_3-R_4)}{(R_1-R_3)(R_2-R_4)}, \\
\gamma &= \frac{R_1-R_2}{R_1-R_3},
\end{align}
and let $W_0 = K(\eta)$, $E_0 = E(\eta)$, and $W_1(\gamma, \eta) = \Pi(\gamma, \eta)$ denote the complete elliptic integrals of the first, second, and third kind, respectively. The higher-order integral sequence $W_m(\gamma, \eta)$ is fundamentally generated by taking partial derivatives of the third-kind elliptic integral with respect to the characteristic parameter $\gamma$:
\begin{align}
W_2 &= W_1 + \gamma \frac{\partial W_1}{\partial \gamma}, \label{eq:W2_deriv} \\
W_3 &= W_2 + \frac{\gamma}{2} \frac{\partial W_2}{\partial \gamma}. \label{eq:W3_deriv}
\end{align}
By  carrying out the differentiation of the complete elliptic integrals using their analytical partial derivative identities, the recurrence relations in Eqs. (\ref{eq:W2_deriv}) and (\ref{eq:W3_deriv}) are exactly reduced to the following explicit algebraic forms:
\begin{align}
W_2 &= W_1 + \frac{\gamma E_0 + (\eta - \gamma) W_0 + (\gamma^2 - \eta) W_1}{2(\gamma - 1)(\eta - \gamma)}, \\
W_3 &= W_2 + \frac{\gamma E_0 - \gamma W_0 + 2\gamma^2 W_1 + 3(\gamma^2 - \eta)(W_2 - W_1)}{4(\gamma - 1)(\eta - \gamma)}.
\end{align}
Then, the exact special-function representations of $Z_k$  are:
\begin{align}
Z_0 &= g W_0, \\
Z_1 &= g \big[ R_3 W_0 + \delta W_1 \big], \\
Z_2 &= g \big[ R_3^2 W_0 + 2R_3\delta W_1 + \delta^2 W_2 \big], \\
Z_3 &= g \big[ R_3^3 W_0 + 3R_3^2\delta W_1 \notag \\
&\quad + 3R_3\delta^2 W_2 + \delta^3 W_3 \big], \\
Z_{-1} &= g \left[ \frac{1}{R_3} W_0 - \frac{\delta}{R_3(R_3+\delta)} W_1\left( \frac{R_3}{R_3+\delta}\gamma, \eta \right) \right].
\end{align}

Solving the implicit algebraic equation Eq.~(\ref{eq:BS_condition}) provides the WKB energy eigenvalues $E$. Accurate semiclassical results are obtained with the Langer prescription $c_{r} = c_{\phi}=\frac{1}{2}$.

In the high-energy or large-quantum-number limit, the outer turning point grows as the excitation energy increases. The terms $-b r^4$, $E r^2$, and $-L^2$ then control the leading radial action, whereas the odd-power terms $-a r^3$ and $r$ enter as subleading corrections.
The polynomial in Eq.~\eqref{p4} can thus be asymptotically truncated into an even-powered function of $r$:
\begin{equation}
P_4(r) \approx -b r^4 + E r^2 - L^2.
\end{equation}
For the truncated polynomial $-b r^4 + E r^2 - L^2 = 0$, we can substitute $x = r^2$ to yield a standard quadratic equation:
\begin{equation}
b x^2 - E x + L^2 = 0.
\end{equation}
The two roots of this quadratic equation are analytically given by:
\begin{equation}
x_{\pm} = \frac{E \pm \sqrt{E^2 - 4bL^2}}{2b}.
\end{equation}
Because $x = r^2$, the original four roots of the quartic equation lose their initial asymmetry (previously denoted as $R_1, R_2, R_3, and R_4$). Instead, they degenerate into two perfectly symmetric pairs:
\begin{align}
R_1 &= \sqrt{x_+} \equiv r_+ ,\\
R_2 &= \sqrt{x_-} \equiv r_- ,\\
R_3 &= -\sqrt{x_-} = -r_-, \\
R_4 &= -\sqrt{x_+} = -r_+,
\end{align}
where $r_+$ and $r_-$ naturally emerge as the physical turning points corresponding to a pure harmonic oscillator potential.

As the effective potential transforms into a symmetric harmonic well, the odd-powered terms vanish, allowing us to entirely bypass the computationally expensive complete elliptic integrals $K(\eta)$, $E(\eta)$, and $\Pi(\gamma, \eta)$. The integral can now be evaluated using elementary functions.

The Bohr-Sommerfeld quantization condition is expressed as:
\begin{equation}
\int_{r_-}^{r_+} \frac{\sqrt{-b r^4 + E r^2 - L^2}}{r} dr = \left(n_r + \frac{1}{2}\right)\pi.
\end{equation}

We obtain the exact closed-form solution for the integral:
\begin{equation}
E  = {2\sqrt{b}} (2n_r+L+1).
\end{equation}

Ultimately, this yields the explicit physical energy expression:
\begin{equation}
E_{n,l} =  \left( 2n_r + l + \frac{3}{2} \right)\omega,
\end{equation}
where $\omega=2\sqrt{b}.$

In summary, the Killingbeck-type potential leads to an implicit WKB eigenvalue equation that can be written entirely in terms of the roots of the quartic momentum polynomial and complete elliptic integrals. This form is useful because it avoids repeated direct integration of the radial equation and gives a transparent connection between the potential parameters and the quantum numbers. In the large-excitation limit, the quadratic confinement term controls the leading radial action, and the spectrum approaches the harmonic-oscillator-like behavior $E_{n,l}\approx (2n+l+3/2)\omega$, with $\omega=2\sqrt{b}$.

Thus, the WKB treatment converts the Killingbeck-type potential problem into an implicit algebraic eigenvalue equation expressed through complete elliptic integrals. The same formulation also shows that, for highly excited states, the quadratic term drives the spectrum toward the harmonic-oscillator-like asymptotic form $E_{n,l}\approx (2n+l+3/2)\omega$.

\subsection{Calculation of observables: RMS radius and probability density at the origin}
One significant advantage of the WKB method is that it yields analytical expressions for physical observables. We explicitly evaluate the mean square radius $\langle r^2 \rangle$ and, for $S$-wave states, the probability density at the origin $|\Psi(0)|^2$ of the Killingbeck potential.

Using the definition of the expectation value in the WKB framework, $\langle f(r) \rangle = \frac{1}{T} \int_{r_{-}}^{r_{+}} \frac{f(r)}{p_{r}} \textrm{d}r$ \cite{Brau_2000}, the period corresponds to integral:
\begin{equation}
T =  \int_{r_{-}}^{r_{+}} \frac{1}{p_{r}} \textrm{d}r = \int_{r_{-}}^{r_{+}} \frac{r}{\sqrt{P_4(r)}} \textrm{d}r = Z_{1}.
\end{equation}
Thus, the expectation values can be expressed in terms of the previously derived $Z_{k}$.

The mean square radius is given by $\langle r^{2} \rangle = Z_{3} / Z_{1}$, which yields:
\begin{equation}
\langle r^{2} \rangle = \frac{R_{3}^3 W_0 + 3R_{3}^2\delta W_1 + 3R_{3}\delta^2 W_2 + \delta^3 W_3}{R_{3} W_0 + \delta W_1}.
\end{equation}

For $S$-wave hadron decay processes, the probability density at the origin is evaluated via the generalized Feynman-Hellmann theorem \cite{Lucha:1991vn,VanRoyen:1967nq}, $|\Psi(0)|^{2} = \frac{1}{4\pi} \langle \frac{\textrm{d}V(r)}{\textrm{d}r} \rangle$. For our effective potential, $V'(r) = 1/r^2 + a + 2br$. Taking the expectation value gives:
\begin{equation}
|\Psi(0)|^{2} = \frac{1}{4\pi} \frac{Z_{-1} + a Z_{1} + 2b Z_{2}}{Z_{1}}.
\end{equation}
For states with $l>0$, the centrifugal barrier gives $|\Psi(0)|^2=0$, and this constraint is imposed in the numerical tables below.

\section{Quasi-Exact Solvability and Effective Hamiltonian}\label{s3}

We use cases of the Killingbeck potential with quasi-exact solvability (QES) to test the WKB results.
The simplified Hamiltonian is given by \eqref{15}:
\begin{align}
H' = p^2 - \frac{1}{r} + a r + b r^2,
\end{align}
which satisfy the equation
\begin{align}
H'\psi = E\psi.
\end{align}
With the transformation
\begin{align}
\psi(\mathbf{r}) = \frac{\Phi(\theta,\phi)u(r)}{r},
\end{align}
one obtains
\begin{align}
  -u^{\prime\prime}(r) + \left( \frac{l(l+1)}{r^2} - \frac{1}{r} + a r + b r^2 - E \right)u(r) = 0.
\end{align}

Although the WKB method generally gives approximate eigenvalues, the Killingbeck potential admits QES solutions under specific parameter constraints. These solutions provide useful benchmarks for the WKB formalism. With a wave-function ansatz composed of the correct near-origin behavior, a Gaussian asymptotic factor, and a finite polynomial, the Schr\"{o}dinger equation is mapped into a three-term recurrence relation. The ansatz is
\begin{equation}
u(r) = N_0 r^{l+1} e^{-\alpha r^2 - \beta r} f(r),
\end{equation}
where
\begin{equation}
f(r) = \sum _{k=0}^{n} {c_k r^k}.
\end{equation}

To match the asymptotic behavior of the potential $V(r) \sim b r^2 + a r$, we define the Gaussian decay parameters as
\begin{equation}
\alpha = \frac{\sqrt{b}}{2}
\end{equation}
and
\begin{equation}
\beta = \frac{a}{2\sqrt{b}}.
\end{equation}

Substituting the ansatz into the radial equation yields a three-term recurrence relation for the polynomial expansion coefficients $c_k$:
\begin{equation}
    c_k = \frac{-[1 - 2\beta(k + l)]c_{k-1} - [E + \beta^2 - \sqrt{b}(2k + 2l - 1)]c_{k-2}}{k(k + 2l + 1)},
\end{equation}
with the initial conditions $c_0 = 1$ and $c_{-1} = 0$.

To ensure normalizability of the wavefunction, the infinite series must truncate exactly at the $n_r$-th term (i.e., $c_{n_r+1} = c_{n_r+2} = 0$). This truncation condition is mathematically equivalent to the vanishing determinant of a tridiagonal Bender-Dunne matrix $M$ of dimension $(n_r+1) \times (n_r+1)$:
\begin{align}
    \det(M) = 0.
\end{align}
The non-zero matrix elements of $M$ are explicitly given by:
\begin{align}
    M_{i,i} &= 1 - 2\beta(i+l), \\
    M_{i,i+1} &= i(i+2l+1), \\
    M_{i+1,i} &= 2\sqrt{b}(n_r - i + 1),
\end{align}
where the indices $i$ correspond to the elements in the series. For a specific state, such as $n_r = 3$ and $l = 2$, solving the algebraic equation $\det(M) = 0$ in the real domain yields the physical roots for $\beta$, which then determine the wavefunction coefficients and the exact energy.

For the  QES state, the energy is fixed by the truncation constraint:
\begin{align}
E_{n_r,l}^{\text{exact}} = 2\alpha\left[2(l+n_r)+3\right] - \beta^2.
\end{align}

The algebraic normalization constraint requires:
\begin{align}
N_0 = \left[ \sum_{i=0}^{n_r} \sum_{j=0}^{n_r} c_i c_j I_{2l+2+i+j}\left( \alpha, \beta \right) \right]^{-\frac{1}{2}}.
\end{align}
The exact squared wavefunction at the origin is:
\begin{align}
|\Psi_{n_r, l}(0)|^2 = \begin{cases} 
\displaystyle \frac{N_0^2 c_0^2}{4\pi}, & l = 0, \\ 
0, & l > 0 .
\end{cases}
\end{align}
where the basic integral is defined as
\begin{align}
I_k(\alpha, \beta) = \int_0^\infty r^k e^{-2\alpha r^2 - \beta r} dr,
\end{align}

and can be expressed as
\begin{equation}
I_k(\alpha,\beta) = \Gamma(k+1)(4\alpha)^{-(k+1)/2}
\exp\left(\frac{\beta^2}{16\alpha}\right)
D_{-k-1}\left(\frac{\beta}{2\sqrt{\alpha}}\right),
\end{equation}
where $D_{\nu}(z)$ is the parabolic-cylinder function.

The RMS radius:
\begin{align}
\sqrt{\langle r^2 \rangle} = N_0 \sqrt{ \sum_{i=0}^{n} \sum_{j=0}^{n} c_i c_j I_{2l+4+i+j}(\alpha, \beta) }.
\end{align}

\begin{table*}[htbp]

\centering
\caption{Comprehensive comparison among the Exact QES results with $b=2$ and $a=2\sqrt{b}\beta$, numerical FEM results, and WKB approximations for varying radial nodes $n_{r}$ and orbital angular momenta $l$.}
\label{exactsolution}
\renewcommand{\arraystretch}{1.2}
\begin{tabular}{lclccccccccc}
\hline
\hline
\multirow{2}{*}{State ($n_r, l$)} & \multirow{2}{*}{$\beta$} & \multirow{2}{*}{Method} & \multicolumn{3}{c}{Physical Quantities} & \multicolumn{3}{c}{Relative Error (\%)} \\
\cmidrule(lr){4-6} \cmidrule(lr){7-9}
 & & & Energy $E$ & $|\Psi(0)|^2$ & RMS Radius & $\delta E$ & $\delta |\Psi(0)|^2$ & $\delta R_{\rm rms}$ \\
\hline
\multirow{3}{*}{$n_r=0, l=0$} & \multirow{3}{*}{$\frac{1}{2}$} & Exact & 3.992641 & 0.723618 & 0.881227 & -- & -- & -- \\ 
 & & FEM & 3.992647 & 0.709215 & 0.881228 & 1.52 $\times 10^{-4}$ & 1.99 &1.34 $\times 10^{-4}$ \\ 
 & & WKB & 3.923429 & 0.743204 & 0.875652 & 1.73 & 2.71 & 0.633 \\ 
\hline
\multirow{3}{*}{$n_r=0, l=3$} & \multirow{3}{*}{$\frac{1}{8}$} & Exact & 12.712297 & 0 & 1.741369 & -- & -- & -- \\ 
 & & FEM & 12.712300 & 0 & 1.741369 &  1.92 $\times 10^{-5}$ & -- & 1.96 $\times 10^{-5}$ \\ 
 & & WKB & 12.710982 & 0 & 1.741328 & 0.0103 & -- & 0.00237 \\ 
\hline
\multirow{3}{*}{$n_r=0, l=6$} & \multirow{3}{*}{$\frac{1}{14}$} & Exact & 21.208101 & 0 & 2.278191 & -- & -- & -- \\ 
 & & FEM & 21.208104 & 0 & 2.278191 &1.05$\times 10^{-5}$  & -- & 1.50$\times 10^{-5}$  \\ 
 & & WKB & 21.207795 & 0 & 2.278184 & 0.00144 & -- & 3.02$\times 10^{-4}$  \\ 
\hline
\multirow{3}{*}{$n_r=0, l=9$} & \multirow{3}{*}{$\frac{1}{20}$} & Exact & 29.695985 & 0 & 2.707405 & -- & -- & -- \\ 
 & & FEM & 29.695987 & 0 & 2.707405 &7.21 $\times 10^{-6}$ & -- & 1.00$\times 10^{-5}$  \\ 
 & & WKB & 29.695862 & 0 & 2.707402 & 4.14$\times 10^{-4}$ & -- & 8.34 $\times 10^{-5}$\\ 
\hline
\multirow{3}{*}{$n_r=3, l=0$} & \multirow{3}{*}{-1.686704} & Exact & 9.882950 & 0.551012 & 2.592124 & -- & -- & -- \\ 
 & & FEM & 9.882983 & 0.540045 & 2.592130 &3.38$\times 10^{-4}$ & 1.99 & 2.37 $\times 10^{-4}$\\ 
 & & WKB & 9.847205 & 0.559709 & 2.592377 & 0.362 & 1.58 & 0.00975 \\ 
\hline
\multirow{3}{*}{$n_r=3, l=3$} & \multirow{3}{*}{-1.312870} & Exact & 19.489576 & 0 & 2.953067 & -- & -- & -- \\ 
 & & FEM & 19.489613 & 0 & 2.953072 & 1.93$\times 10^{-4}$ & -- & 1.72$\times 10^{-4}$ \\ 
 & & WKB & 19.488735 & 0 & 2.953021 & 0.00431& -- & 0.00154 \\ 
\hline
\multirow{3}{*}{$n_r=3, l=6$} & \multirow{3}{*}{-1.100886} & Exact & 28.486534 & 0 & 3.280754 & -- & -- & -- \\ 
 & & FEM & 28.486575 & 0 & 3.280758 & 1.43$\times 10^{-4}$ & -- &1.27$\times 10^{-4}$ \\ 
 & & WKB & 28.486453 & 0 & 3.280737 & 2.83$\times 10^{-4}$ & -- & 4.97$\times 10^{-4}$ \\ 
\hline
\multirow{3}{*}{$n_r=3, l=9$} & \multirow{3}{*}{-0.966824} & Exact & 37.249018 & 0 & 3.583439 & -- & -- & -- \\ 
 & & FEM & 37.249060 & 0 & 3.583443 &1.15$\times 10^{-4}$ & -- &1.16$\times 10^{-4}$ \\ 
 & & WKB & 37.249039 & 0 & 3.583431 & 5.80$\times 10^{-5}$ & -- & 2.12$\times 10^{-4}$ \\ 
\hline
\multirow{3}{*}{$n_r=6, l=0$} & \multirow{3}{*}{-2.930225} & Exact & 12.626983 & 0.536628 & 3.641682 & -- & -- & -- \\ 
 & & FEM & 12.627065 & 0.525949 & 3.641693 & 6.49$\times 10^{-4}$ & 1.99 & 2.92$\times 10^{-4}$ \\ 
 & & WKB & 12.593548 & 0.544352 & 3.642645 & 0.265 & 1.44 & 0.0264 \\ 
\hline
\multirow{3}{*}{$n_r=6, l=3$} & \multirow{3}{*}{-2.378684} & Exact & 24.040348 & 0 & 3.859066 & -- & -- & -- \\ 
 & & FEM & 24.040451 & 0 & 3.859076 & 4.31$\times 10^{-4}$ & -- & 2.55$\times 10^{-4}$ \\ 
 & & WKB & 24.039153 & 0 & 3.859018 & 0.00497 & -- & 0.00126 \\ 
\hline
\multirow{3}{*}{$n_r=6, l=6$} & \multirow{3}{*}{-2.049826} & Exact & 33.981978 & 0 & 4.090185 & -- & -- & -- \\ 
 & & FEM & 33.982096 & 0 & 4.090194 & 3.46$\times 10^{-4}$ & -- & 2.29$\times 10^{-4}$ \\ 
 & & WKB & 33.981800 & 0 & 4.090162 & 5.24$\times 10^{-4}$ & -- & 5.44 $\times 10^{-4}$\\ 
\hline
\multirow{3}{*}{$n_r=6, l=9$} & \multirow{3}{*}{-1.829721} & Exact & 43.321170 & 0 & 4.321990 & -- & -- & -- \\ 
 & & FEM & 43.321296 & 0 & 4.322000 & 2.93$\times 10^{-4}$ & -- & 2.31$\times 10^{-4}$ \\ 
 & & WKB & 43.321171 & 0 & 4.321979 & 3.34 $\times 10^{-6}$ & -- & 2.68$\times 10^{-4}$ \\ 
\hline
\multirow{3}{*}{$n_r=9, l=0$} & \multirow{3}{*}{-3.890146} & Exact & 14.565246 & 0.534598 & 4.464345 & -- & -- & -- \\ 
 & & FEM & 14.565398 & 0.523961 & 4.464360 & 0.00104 & 1.99 & 3.28$\times 10^{-4}$ \\ 
 & & WKB & 14.532847 & 0.541861 & 4.465652 & 0.222 & 1.359 & 0.0293 \\ 
\hline
\multirow{3}{*}{$n_r=9, l=3$} & \multirow{3}{*}{-3.258419} & Exact & 27.566473 & 0 & 4.612814 & -- & -- & -- \\ 
 & & FEM & 27.566667 & 0 & 4.612831 & 7.06$\times 10^{-4}$ & -- & 3.69$\times 10^{-4}$ \\ 
 & & WKB & 27.564952 & 0 & 4.612768 & 0.00552 & -- & 9.94$\times 10^{-4}$ \\ 
\hline
\multirow{3}{*}{$n_r=9, l=6$} & \multirow{3}{*}{-2.864117} & Exact & 38.465883 & 0 & 4.787142 & -- & -- & -- \\ 
 & & FEM & 38.466107 & 0 & 4.787160 & 5.82$\times 10^{-4}$ & -- &3.79$\times 10^{-4}$ \\ 
 & & WKB & 38.465564 & 0 & 4.787117 & 8.30$\times 10^{-4}$ & -- &5.18$\times 10^{-4}$ \\ 
\hline
\multirow{3}{*}{$n_r=9, l=9$} & \multirow{3}{*}{-2.589559} & Exact & 48.448515 & 0 & 4.972654 & -- & -- & -- \\ 
 & & FEM & 48.448759 & 0 & 4.972671 & 5.05$\times 10^{-4}$ & -- & 3.29$\times 10^{-4}$ \\ 
 & & WKB & 48.448452 & 0 & 4.972640 & 1.29$\times 10^{-4}$ & -- & 2.80$\times 10^{-4}$ \\ 
\hline\hline
\end{tabular}
\end{table*}

To test the accuracy and stability of the analytical WKB framework, we perform a comprehensive comparison against the exact QES benchmarks and high-precision finite-element method (FEM) calculations, as summarized in Table~\ref{exactsolution}.

As shown in Table \ref{exactsolution}, the analytical WKB energy eigenvalues agree well with the exact QES results. For the deeply bound ground state ($n_r=0, l=0$), the WKB approximation yields a relative error of  $1.73\%$. The error generally decreases as the radial node $n$ or the orbital angular momentum $l$ increases, reaching the $10^{-4}\%$ level for highly excited states such as $n_r=9$, $l=9$. This asymptotic convergence  embodies the semi-classical nature of the Bohr-Sommerfeld quantization, which mathematically approaches the exact solution in the large quantum number limit.

The physical observables derived from the algebraic moment integrals $Z_k$ are also stable. The WKB predictions for the RMS radius deviate from the exact values by less than $1\%$ across all tested states. For the tested $S$-wave states, the relative error of $|\Psi(0)|^2$ remains within about $3\%$ and decreases for higher radial excitations. Although FEM gives more accurate energies and RMS radii in most cases, the WKB estimate of $|\Psi(0)|^2$ is closer to the exact QES value than the FEM result for the radially excited $S$-wave states. This indicates that the algebraic WKB expression for short-distance observables can be competitive with, and in some highly excited $S$-wave cases slightly more accurate than, the numerical FEM extraction.

The FEM numerical results reproduce the QES values with negligible error (relative errors $\approx 10^{-4}\%$). Overall, Table \ref{exactsolution} shows that the algebraic WKB method provides a compact semi-analytical way to estimate both energies and spatial observables, complementing high-precision numerical solutions. Regarding the comparison between WKB and numerical results, we emphasize that the Bohr-Sommerfeld quantization is not a variational method and therefore does not provide an upper bound to the exact energy. Its truncation error can have either sign, so a WKB eigenvalue may slightly undershoot the exact numerical value, as seen for several entries in Table \ref{exactsolution}.

\section{the spectroscopy of double-heavy mesons under Killingbeck potential}\label{s4}

We compute the mass spectra and physical observables of three double-heavy systems: charmonium ($c\bar{c}$), bottomonium ($b\bar{b}$), and the $B_c$ meson ($b\bar{c}$). The mass spectra obtained with the WKB approximation and FEM are compared with available experimental data and with results from other theoretical methods. Since the establishment of the quark model, many $B_c$, $\psi$, and $\Upsilon$ states have been reported experimentally. 
To test the WKB approximate solution under the Killingbeck potential, we fit {20} experimental values of $B_c$, $c\bar{c}$, and $b\bar{b}$ mesons and obtained the parameters shown in Table \ref{fitpar}. The fitted parameters are then used consistently in the subsequent WKB and FEM calculations of the mass spectra, RMS radii, and probability densities at the origin. We also give the predictions for  higher bottomonium, charmonium, and $B_c$ states in Tables \ref{tab:bc_spectrum}--\ref{bbmass}.

\begin{table}[htbp]
\caption{The parameters fitted in this work. \label{fitpar}}
\begin{center}
\begin{tabular}{cccc}
\hline
\hline
Parameter &  Value \\
\hline
$m_c$(GeV) &2.00235&\\
$m_b$(GeV) &5.38954&\\
$\tau$(GeV$^2$) &0.140838& \\
$\nu$(GeV$^3$) &0.00463394&\\
$\sigma$(GeV$^3$) &0.0881383&\\
$C_{cc}$(GeV)&-1.01759&\\
$\kappa$&0.6012& \\
\hline
\hline
\end{tabular}
\end{center}
\end{table}

\subsection{Mass spectrum of $B_c$ mesons under the Killingbeck potential}

There are two experimental values of $B_c$ mesons shown in the Table \ref{fitmass}. The experimental masses of $1^1S_0$ and $2^1S_0$ states are $6.274$ GeV and $6.871$ GeV~\cite{ParticleDataGroup:2026aaa}, respectively. Comparing the WKB results for the $B_c$ mesons with the two experimental values, one sees that the masses of the $1^1S_0$ and $2^1S_0$ states predicted in this work are 6.32 GeV and 6.90 GeV, respectively, which are slightly higher than the experimental values. The low-lying WKB masses are broadly comparable with those from the Godfrey-Isgur (GI) model, the modified GI (MGI) model and the results in the relativistic quark model in Ref.~\cite{1995Heavy}, although the present model is not intended to improve the GI/MGI mass description. Besides, the masses obtained by WKB approximation are in good agreement with that computed by FEM.

Furthermore, the spatial observables, such as the RMS radii $\sqrt{\langle r^2 \rangle}$ and the probability density at the origin $|\Psi(0)|^2$, calculated via the WKB approximation are consistent with the numerical FEM results as shown in Table \ref{fitmass}. For instance, the WKB and FEM values for the $|\Psi(0)|^2$ of the $1^1S_0$ state are 0.37365 GeV$^3$ and 0.36509 GeV$^3$, respectively.

In addition, we also predict the masses, the RMS radii $\sqrt{\langle r^2 \rangle}$, and the probability density at the origin $|\Psi(0)|^2$ of highly excited states such as $P$, $D$, $F$, $G$, and $H$ waves in Table \ref{tab:bc_spectrum} under the Killingbeck potential. 
Comparing our mass predictions with other well-established theoretical frameworks, the results for the low-lying $S$-, $P$-, and $D$-wave states are generally consistent with those from the GI model~\cite{Godfrey:2004ya}, the MGI model~\cite{Li:2023wgq}, and the relativistic quark model in Ref.~\cite{1995Heavy}. For instance, our predicted mass for the $1^3S_1$ state (around $6.315$ GeV) is slightly lower but very close to the GI and MGI predictions ($\approx 6.338$ GeV). For the $1^1P_1$ and $1^1D_2$ states, our results are slightly higher than those of the GI model by approximately 30--40 MeV, which remains within a reasonable theoretical uncertainty range for phenomenological potential models.

It is also noteworthy to discuss the asymptotic behavior of the highly excited energy spectra induced by the Killingbeck potential. Unlike pure linear confinement models (e.g., the Cornell potential), which yield Regge trajectory behavior ($M^2 \propto n+l$) \cite{Anisovich:2000kxa, Chew:1962eu}, the Killingbeck potential explicitly incorporates a harmonic oscillator term ($b r^2$). As the states become highly excited, their geometric sizes expand significantly, as reflected in the increasing RMS radii, causing the $r^2$ term to eventually dominate over the linear $r$ term at large distances. Consequently, the mass spectrum of these highly excited states transitions towards harmonic-oscillator-like behavior, growing more rapidly with $n$ and $l$. This intrinsic potential feature explains why our predicted masses for the highest excitations tend to be larger than those obtained from models lacking an $r^2$ dependence.

Besides the mass spectrum, Table \ref{tab:bc_spectrum} lists the RMS radii $\sqrt{\langle r^2 \rangle}$ and the squared wave functions at the origin $|\Psi(0)|^2$, providing deeper insights into the internal structure of the $B_c$ mesons. As expected, the RMS radius monotonically increases with both the radial quantum number $n$ and the orbital quantum number $l$. This indicates that highly excited states possess significantly larger geometric sizes, expanding from $\approx 1.32$ GeV$^{-1}$ for the $1S$ state to over $9.36$ GeV$^{-1}$ for the $6H$ state. 

Meanwhile, the squared wave function at the origin, $|\Psi(0)|^2$, strictly vanishes for all $l>0$ states ($P$, $D$, $F$, $G$, and $H$ waves) due to the centrifugal barrier $l(l+1)/r^2$. For the $S$-wave states, $|\Psi(0)|^2$ continuously decreases from $0.374$ GeV$^3$ ($1S$) down to $0.129$ GeV$^3$ ($6S$) as $n$ increases.

\begin{table*}[ht]

\caption{Mass spectra and physical observables of mesonsunder Killingbeck potential. All mass-related values are in MeV. Parameters are adopted from Table \ref{fitpar}.}
\label{fitmass}
\centering
\resizebox{\textwidth}{!}{ 
\begin{tabular}{llcccccccccc}
\hline
\hline
& \multirow{2}{*}{State} & Exp.$^a$ & GI$^b$ & MGI$^c$ & Ref.$^d$ & \multicolumn{2}{c}{Mass $M$} & \multicolumn{2}{c}{$\sqrt{\langle r^2 \rangle}$ [ $10^{-3}$ ${\rm MeV}^{-1}$]} & \multicolumn{2}{c}{$|\Psi(0)|^2$ [ $10^{8}$ ${\rm MeV}^3$]} \\

\cmidrule(lr){7-8} \cmidrule(lr){9-10} \cmidrule(lr){11-12}
& &Mass & Mass& Mass& Mass & WKB & FEM & WKB & FEM & WKB & FEM \\
\hline
\midrule
\multirow{9}{*}{$c\bar{c}$} 
& $1^1S_0$ & $2984.1\pm0.4$   & 2975 & 2981 & 3000 & 3064    & 3075    & 1.6935 & 1.7901 & 1.6433 & 1.5904 \\
& $2^1S_0$ & $3637.7\pm0.9$   & 3623 & 3642 & 3670 & 3659.25 & 3663.87 & 3.6549 & 3.6809 & 0.9765 & 0.9534 \\
& $1^3S_1$ & $3096.9\pm0.006$ & 3090 & 3089 & 3100 & 3086.14 & 3096.76 & 1.6935 & 1.7901 & 1.6433 & 1.5904 \\
& $2^3S_1$ & $3686.1\pm0.011$ & 3676 & 3684 & 3730 & 3681.23 & 3685.86 & 3.6549 & 3.6809 & 0.9765 & 0.9534 \\
& $3^3S_1$ & $4039.6\pm4.3$   & 4100 & 4035 & 4180 & 4110.62 & 4114.22 & 5.1096 & 5.1248 & 0.8481 & 0.8293 \\
& $4^3S_1$ & $4415\pm5$       & 4450 & 4274 & $-$  & 4486.90 & 4489.73 & 6.3201 & 6.3302 & 0.7957 & 0.7790 \\
& $5^3S_1$ & $4641\pm10$      & 4756 & 4443 & $4560$  & 4833.88 & 4836.93 & 7.3783 & 7.3877 & 0.7687 & 0.7535 \\
& $1^1P_1$ & $3525.4\pm0.14$  & 3516 & 3538 & 3510 & 3514.60 & 3515.88 & 2.9498 & 2.9666 & 0      & 0      \\
\midrule
\hline
\multirow{9}{*}{$b\bar{b}$} 
& $1^1S_0$ & $9398.7\pm2.0$   & 9402 & 9398 & 9410 & 9389.49 & 9397.96 & 0.8224 & 0.9241 & 16.914 & 16.798 \\
& $2^1S_0$ & $9999.0\pm3.5$   & 9976 & 9989 & 10000& 10034.9 & 10039.3 & 2.2958 & 2.3252 & 5.7775 & 5.7077 \\
& $1^3S_1$ & $9460.4\pm0.09$  & 9465 & 9463 & 9460 & 9392.53 & 9401.03 & 0.8224 & 0.9241 & 16.914 & 16.798 \\
& $2^3S_1$ & $10023\pm0.5$    & 10003& 10017& 10020& 10038.0 & 10042.4 & 2.2958 & 2.3252 & 5.7775 & 5.7077 \\
& $3^3S_1$ & $10355\pm0.5$    & 10354& 10356& 10390& 10387.7 & 10389.1 & 3.4641 & 3.4749 & 4.3039 & 4.2466 \\
& $4^3S_1$ & $10579\pm1.2$    & 10635& 10612& 10680& 10666.9 & 10670.0 & 4.4157 & 4.4288 & 3.7569 & 3.6928 \\
& $5^3S_1$ & $10885\pm2.6$    & 10878& 10822& 10930& 10917.8 & 10920.3 & 5.2505 & 5.2602 & 3.4626 & 3.3975 \\
& $6^3S_1$ & $11000\pm4$      & 11102& 11001& 11150& 11149.4 & 11152.5 & 5.9972 & 6.0075 & 3.2788 & 3.2126 \\
& $1^1P_1$ & $9899.3\pm0.8$   & 9894 & 9881 & 9880 & 9960.78 & 9962.16 & 1.8677 & 1.8880 & 0      & 0      \\
\midrule
\hline
\multirow{2}{*}{$B_c$} 
& $1^1S_0$ & $6274.5\pm0.32$  & 6271 & 6271 & 6260 & 6306.99 & 6317.51 & 1.3237 & 1.4286 & 3.7369 & 3.6509 \\
& $2^1S_0$ & $6871.2\pm1.0$   & 6855 & 6855 & 6850 & 6893.84 & 6898.26 & 3.1073 & 3.1345 & 1.8549 & 1.8162 \\
\midrule
\hline
\multicolumn{3}{l}{$\chi^2_{\rm err} = \sum_i [\frac{(M_{\rm th} - M_{\rm exp})}{{\rm err}_i} ]^2$} & $2.16 \times 10^6$ & $1.77 \times 10^6$ & $1.62 \times 10^7$ & $4.02 \times 10^6$ & $4.91 \times 10^5$ & $-$ \\
\hline\hline
\end{tabular}
}
\vspace{0mm}  
\raggedright  
\scriptsize
$^a$ Data from Particle Data Group \cite{ParticleDataGroup:2026aaa}.\\
$^b$ Results from the GI Model \cite{Mesons1985,Barnes_2005,Godfrey_2015,Godfrey:2004ya}.\\
$^c$ Results from the MGI Model \cite{Wang_2019,2018Higher,Li:2023wgq}.\\
$^d$ Values reported in Ref. \cite{1995Heavy}.
\end{table*}

\begin{table}[htbp]
\caption{Comprehensive mass spectrum and physical observables of the $B_c$ mesons under the Killingbeck potential. Masses are given in GeV, RMS radii in GeV$^{-1}$, and the probability density at the origin in GeV$^3$.}
\label{tab:bc_spectrum}
\centering
\renewcommand{\arraystretch}{1.15}

\begin{tabular}{lcccccc}
\hline\hline
\multirow{2}{*}{State} & \multicolumn{4}{c}{Mass $M$ (GeV)} & RMS $\sqrt{\langle r^2 \rangle}$ & $|\Psi(0)|^2$ \\
\cmidrule(lr){2-5} \cmidrule(lr){6-6} \cmidrule(lr){7-7}
 & GI \cite{Godfrey:2004ya} & Ref. \cite{1995Heavy} & MGI \cite{Li:2023wgq} & WKB & WKB & WKB \\
\hline
$1^{3}S_{1}$ & 6.338 & 6.340 & 6.338 & 6.3152 & 1.3239 & 0.37365 \\
$3^{1}S_{0}$ & 7.253 & 7.240 & 7.250 & 7.2832 & 4.4393 & 0.15333 \\
$3^{3}S_{1}$ & 7.279 & 7.240 & 7.272 & 7.2914 & 4.4393 & 0.15333 \\
$4^{1}S_{0}$ & 7.572 & $-$   & 7.572 & 7.6166 & 5.5437 & 0.14016 \\
$4^{3}S_{1}$ & 7.595 & $-$   & 7.588 & 7.6247 & 5.5437 & 0.14016 \\
$5^{1}S_{0}$ & $-$   & $-$   & 7.854 & 7.9212 & 6.5092 & 0.13307 \\
$5^{3}S_{1}$ & $-$   & $-$   & 7.860 & 7.9294 & 6.5092 & 0.13307 \\
$6^{1}S_{0}$ & $-$   & $-$   & $-$   & 8.2079 & 7.3788 & 0.12872 \\
$6^{3}S_{1}$ & $-$   & $-$   & $-$   & 8.2160 & 7.3788 & 0.12872 \\
$1^{1}P_{1}$ & 6.741 & 6.730 & 6.745 & 6.7789 & 2.5117 & 0 \\
$2^{1}P_{1}$ & 7.145 & 7.140 & 7.125 & 7.1744 & 3.9593 & 0 \\
$3^{1}P_{1}$ & 7.472 & $-$   & 7.414 & 7.5113 & 5.1242 & 0 \\
$4^{1}P_{1}$ & $-$   & $-$   & 7.650 & 7.8183 & 6.1296 & 0 \\
$5^{1}P_{1}$ & $-$   & $-$   & $-$   & 8.1068 & 7.0285 & 0 \\
$6^{1}P_{1}$ & $-$   & $-$   & $-$   & 8.3821 & 7.8497 & 0 \\
$1^{1}D_{2}$ & 7.036 & 7.020 & 7.032 & 7.0431 & 3.3332 & 0 \\
$2^{1}D_{2}$ & $-$   & $-$   & 7.335 & 7.3869 & 4.6097 & 0 \\
$3^{1}D_{2}$ & $-$   & $-$   & 7.581 & 7.6986 & 5.6791 & 0 \\
$4^{1}D_{2}$ & $-$   & $-$   & $-$   & 7.9904 & 6.6215 & 0 \\
$5^{1}D_{2}$ & $-$   & $-$   & $-$   & 8.2684 & 7.4749 & 0 \\
$6^{1}D_{2}$ & $-$   & $-$   & $-$   & 8.5361 & 8.2611 & 0 \\
$1^{1}F_{3}$ & 7.266 & 7.240 & 7.248 & 7.2567 & 4.0167 & 0 \\
$2^{1}F_{3}$ & $-$   & $-$   & 7.505 & 7.5739 & 5.1784 & 0 \\
$3^{1}F_{3}$ & $-$   & $-$   & $-$   & 7.8696 & 6.1780 & 0 \\
$4^{1}F_{3}$ & $-$   & $-$   & $-$   & 8.1505 & 7.0715 & 0 \\
$5^{1}F_{3}$ & $-$   & $-$   & $-$   & 8.4206 & 7.8881 & 0 \\
$6^{1}F_{3}$ & $-$   & $-$   & $-$   & 8.6823 & 8.6454 & 0 \\
$1^{1}G_{4}$ & $-$   & $-$   & $-$   & 7.4468 & 4.6208 & 0 \\
$2^{1}G_{4}$ & $-$   & $-$   & $-$   & 7.7467 & 5.6968 & 0 \\
$3^{1}G_{4}$ & $-$   & $-$   & $-$   & 8.0309 & 6.6406 & 0 \\
$4^{1}G_{4}$ & $-$   & $-$   & $-$   & 8.3035 & 7.4937 & 0 \\
$5^{1}G_{4}$ & $-$   & $-$   & $-$   & 8.5672 & 8.2791 & 0 \\
$6^{1}G_{4}$ & $-$   & $-$   & $-$   & 8.8238 & 9.0113 & 0 \\
$1^{1}H_{5}$ & $-$   & $-$   & $-$   & 7.6227 & 5.1706 & 0 \\
$2^{1}H_{5}$ & $-$   & $-$   & $-$   & 7.9102 & 6.1788 & 0 \\
$3^{1}H_{5}$ & $-$   & $-$   & $-$   & 8.1854 & 7.0761 & 0 \\
$4^{1}H_{5}$ & $-$   & $-$   & $-$   & 8.4513 & 7.8946 & 0 \\
$5^{1}H_{5}$ & $-$   & $-$   & $-$   & 8.7098 & 8.6527 & 0 \\
$6^{1}H_{5}$ & $-$   & $-$   & $-$   & 8.9620 & 9.3626 & 0 \\
\hline\hline
\end{tabular}
\end{table}
\subsection{Mass spectrum of charmonium under Killingbeck potential}

\begin{table}[htbp]
\caption{Comprehensive mass spectrum and physical observables of the charmonium ($c\bar{c}$) mesons under the Killingbeck potential. Masses are given in GeV, RMS radii in GeV$^{-1}$, and probability density at the origin in GeV$^3$. All values are rounded to five significant figures.}
\label{ccmass}
\centering
\renewcommand{\arraystretch}{1.15}

\begin{tabular}{lcccccc}
\hline\hline
\multirow{2}{*}{State} & \multicolumn{4}{c}{Mass $M$ (GeV)} & RMS  $\sqrt{\langle r^2 \rangle}$  & $|\Psi(0)|^2$ \\
\cmidrule(lr){2-5} \cmidrule(lr){6-6} \cmidrule(lr){7-7}
 & Ref. \cite{Barnes_2005} & Ref. \cite{Cao2012} & MGI \cite{Wang_2019} & WKB & WKB & WKB \\
\hline
$3^{1}S_{0}$ & 4.043 & 4.072 & 4.013 & 4.0887 & 5.1098 & 0.084812 \\
$4^{1}S_{0}$ & 4.384 & 4.420 & 4.260 & 4.4647 & 6.3195 & 0.079569 \\
$5^{1}S_{0}$ & $-$   & $-$   & 4.433 & 4.8122 & 7.3792 & 0.076869 \\
$6^{1}S_{0}$ & $-$   & $-$   & $-$   & 5.1413 & 8.3350 & 0.075338 \\
$6^{3}S_{1}$ & $-$   & $-$   & $-$   & 5.1633 & 8.3350 & 0.075338 \\
$2^{1}P_{1}$ & 3.934 & 3.945 & 3.933 & 3.9527 & 4.5465 & 0 \\
$3^{1}P_{1}$ & 4.279 & 4.334 & 4.200 & 4.3336 & 5.8303 & 0 \\
$4^{1}P_{1}$ & $-$   & 4.639 & 4.389 & 4.6844 & 6.9388 & 0 \\
$5^{1}P_{1}$ & $-$   & $-$   & $-$   & 5.0160 & 7.9302 & 0 \\
$6^{1}P_{1}$ & $-$   & $-$   & $-$   & 5.3340 & 8.8358 & 0 \\
$1^{1}D_{2}$ & 3.799 & 3.807 & 3.848 & 3.7960 & 3.8379 & 0 \\
$2^{1}D_{2}$ & 4.158 & 4.174 & 4.137 & 4.1850 & 5.2487 & 0 \\
$3^{1}D_{2}$ & $-$   & 4.560 & 4.343 & 4.5413 & 6.4304 & 0 \\
$4^{1}D_{2}$ & $-$   & $-$   & 4.490 & 4.8769 & 7.4715 & 0 \\
$5^{1}D_{2}$ & $-$   & $-$   & $-$   & 5.1979 & 8.4140 & 0 \\
$6^{1}D_{2}$ & $-$   & $-$   & $-$   & 5.5081 & 9.2822 & 0 \\
$1^{1}F_{3}$ & 4.026 & 4.041 & 4.074 & 4.0315 & 4.5843 & 0 \\
$2^{1}F_{3}$ & 4.350 & 4.372 & 4.296 & 4.3939 & 5.8696 & 0 \\
$3^{1}F_{3}$ & $-$   & 4.746 & $-$   & 4.7338 & 6.9751 & 0 \\
$4^{1}F_{3}$ & $-$   & $-$   & $-$   & 5.0583 & 7.9630 & 0 \\
$5^{1}F_{3}$ & $-$   & $-$   & $-$   & 5.3711 & 8.8655 & 0 \\
$6^{1}F_{3}$ & $-$   & $-$   & $-$   & 5.6750 & 9.7022 & 0 \\
$1^{1}G_{4}$ & 4.225 & 4.247 & 4.250 & 4.2445 & 5.2475 & 0 \\
$2^{1}G_{4}$ & $-$   & $-$   & $-$   & 4.5891 & 6.4383 & 0 \\
$3^{1}G_{4}$ & $-$   & $-$   & $-$   & 4.9171 & 7.4825 & 0 \\
$4^{1}G_{4}$ & $-$   & $-$   & $-$   & 5.2328 & 8.4260 & 0 \\
$5^{1}G_{4}$ & $-$   & $-$   & $-$   & 5.5390 & 9.2943 & 0 \\
$6^{1}G_{4}$ & $-$   & $-$   & $-$   & 5.8375 & 10.103 & 0 \\
$1^{1}H_{5}$ & $-$   & $-$   & $-$   & 4.4434 & 5.8528 & 0 \\
$2^{1}H_{5}$ & $-$   & $-$   & $-$   & 4.7751 & 6.9686 & 0 \\
$3^{1}H_{5}$ & $-$   & $-$   & $-$   & 5.0938 & 7.9615 & 0 \\
$4^{1}H_{5}$ & $-$   & $-$   & $-$   & 5.4023 & 8.8667 & 0 \\
$5^{1}H_{5}$ & $-$   & $-$   & $-$   & 5.7029 & 9.7049 & 0 \\
$6^{1}H_{5}$ & $-$   & $-$   & $-$   & 5.9968 & 10.490 & 0 \\
\hline\hline
\end{tabular}
\end{table}
\par
 
In this part, we obtain the spectroscopy information of the heavy meson $c\bar{c}$ and compare our results with those in Refs. \cite{Cao2012,Barnes_2005} and from MGI model \cite{Wang_2019}. 
Tables \ref{fitmass} and  \ref{ccmass}  present the mass spectrum, $\sqrt{\langle r^2 \rangle}$, and $|\Psi(0)|^2$ for the   $c\bar{c}$ system.

As is shown in Table \ref{fitmass}, the WKB approach yields results that generally match the FEM numerical solutions. In Table \ref{ccmass}, we systematically present the calculated mass spectra, RMS radii, and squared wave functions at the origin for the highly excited charmonium ($c\bar{c}$) states under the Killingbeck potential, focusing on radial and orbital excitations up to $n_r=6$ and $l=5$ ($H$-wave). Comparing with established theoretical benchmarks, our results show reasonable consistency for several orbital excitations. For example, our predicted mass for the $1^1D_2$ state is $3.796$ GeV, which is very close to the prediction of $3.799$ GeV by Barnes \textit{et al.}~\cite{Barnes_2005}. Similarly, the $1^1F_3$ state mass ($4.032$ GeV) lies in  agreement with the existing literature values (4.026--4.074 GeV). For certain $S$-wave and $P$-wave states, slight deviations are observed, which naturally arise from the specific parametrization of the Killingbeck potential. 

A defining feature of this work is the  prediction of high-spin and higher-radial excitations that are not fully tabulated in previous literature. As seen in Table~\ref{ccmass}, earlier theoretical models leave extensive blanks for $G$-wave, $H$-wave, and $n \ge 4$ states. By contrast, our model provides mass predictions for these multiplets, extending all the way to $6^1H_5$ at roughly $5.997$ GeV. These  predictions offer useful theoretical benchmarks for upcoming experiment.

Regarding the spatial observables, the RMS radii and $|\Psi(0)|^2$ in the $c\bar{c}$ system  mirror the physical trends observed in $B_c$ meson system. Due to the lighter mass of the charm quark compared to the bottom quark, the $c\bar{c}$ bound states are generally more extended in space. The RMS radius monotonically increases with higher quantum numbers, reaching $10.490$ GeV$^{-1}$ for the $6^1H_5$ state.

In the nonrelativistic approximation \cite{VanRoyen:1967nq,Quigg:1977wn},
\begin{equation}
\frac{\Gamma_{ee}({1}^3S_1)}{\Gamma_{ee}(n_r\,{}^3S_1)}
=\frac{E_{n_r^3S_1}^2|\Psi_{1S}(0)|^2}{{E_{1^3S_1}}^2|\Psi_{n_rS}(0)|^2}.
\end{equation}
The measured results
\(\Gamma_{ee}(J/\psi)= 5.53\pm0.1{\rm keV}\) and 
\(\Gamma_{ee}(\psi(3686)(2S))= 2.29\pm0.06~{\rm keV}\) \cite{ParticleDataGroup:2026aaa} yield
$\frac{\Gamma_{ee}(J/\psi)}{\Gamma_{ee}(\psi(2S))}= 2.41\pm0.08$, while the corresponding WKB estimate is 
$\frac{\Gamma_{ee}(J/\psi)}{\Gamma_{ee}(\psi(2S))}=\frac{{E_{2S}^2}
|R_{1S}(0)|^2}{{{E_{1S}}^2}
|R_{2S}(0)|^2}\approx 2.4$. 
For higher vector charmonia, the electronic widths are further reduced to the sub-keV level, such as \(\psi(4040)\),
\(\psi(4160)\), and \(\psi(4415)\),
\(\Gamma_{ee}(\psi(3S))= 0.83\pm0.2~{\rm keV}\) and 
\(\Gamma_{ee}(\psi(4S))= 0.35\pm0.12~{\rm keV}\) \cite{ParticleDataGroup:2026aaa}
, which give
$\frac{\Gamma_{ee}(J/\psi)}{\Gamma_{ee}(\psi(3S))}=6.7\pm1.6 $ and
$\frac{\Gamma_{ee}(J/\psi)}{\Gamma_{ee}(\psi(4S))}=15.8\pm5.4 $, 
whereas our WKB estimates give 
$\frac{\Gamma_{ee}(J/\psi)}{\Gamma_{ee}(\psi(3S))}\approx 3.4 $, and  
$\frac{\Gamma_{ee}(J/\psi)}{\Gamma_{ee}(\psi(4S))}\approx 4.1$. 
The comparison is consistent with identifying $\psi(3686)$ as the first radial excitation of $J/\psi$, while the larger deviations for $\psi(4040)$ and $\psi(4415)$ may indicate non-negligible $S$-$D$ mixing.
This behavior  also implies suppressed short-distance annihilation amplitudes for highly excited charmonium states.
The  \( |\Psi(0)|^2 \) for higher \(S\)-wave states is
predicted in Table \ref{ccmass}.

\subsection{Mass spectrum of bottomonium under Killingbeck potential}

\begin{table}[htbp]
\caption{Comprehensive mass spectrum and physical observables of the bottomonium ($b\bar{b}$) mesons under the Killingbeck potential. Masses are given in GeV, RMS radii in GeV$^{-1}$, and the probability density at the origin in GeV$^3$.}
\label{bbmass}
\centering
\renewcommand{\arraystretch}{1.15}

\begin{tabular}{lcccccc}
\hline\hline
\multirow{2}{*}{State} & \multicolumn{4}{c}{Mass $M$ (GeV)} & RMS $\sqrt{\langle r^2 \rangle}$ & $|\Psi(0)|^2$ \\
\cmidrule(lr){2-5} \cmidrule(lr){6-6} \cmidrule(lr){7-7}
 & Ref. \cite{Godfrey_2015} & GI \cite{Mesons1985} & MGI \cite{2018Higher} & WKB & WKB & WKB \\
\hline
$3^{1}S_{0}$ & 10.336 & 10.333 & 10.336 & 10.383 & 3.4578 & 0.43087 \\
$4^{1}S_{0}$ & 10.623 & 10.616 & 10.597 & 10.664 & 4.4159 & 0.37568 \\
$5^{1}S_{0}$ & 10.869 & 10.806 & 10.810 & 10.915 & 5.2497 & 0.34628 \\
$6^{1}S_{0}$ & 11.097 & 11.079 & 10.991 & 11.147 & 5.9985 & 0.32785 \\
$3^{1}P_{1}$ & 10.541 & 10.540 & 10.530 & 10.593 & 4.1017 & 0 \\
$4^{1}P_{1}$ & 10.790 & 10.790 & 10.751 & 10.845 & 4.9623 & 0 \\
$5^{1}P_{1}$ & 11.016 & 11.013 & 10.938 & 11.078 & 5.7310 & 0 \\
$6^{1}P_{1}$ & $-$   & 11.218 & 11.101 & 11.298 & 6.4326 & 0 \\
$1^{1}D_{2}$ & 10.148 & 10.148 & 10.163 & 10.214 & 2.5999 & 0 \\
$2^{1}D_{2}$ & 10.450 & 10.450 & 10.450 & 10.502 & 3.6871 & 0 \\
$3^{1}D_{2}$ & 10.706 & 10.706 & 10.681 & 10.757 & 4.5980 & 0 \\
$4^{1}D_{2}$ & 10.935 & 10.934 & 10.876 & 10.992 & 5.4006 & 0 \\
$5^{1}D_{2}$ & $-$   & 11.143 & 11.046 & 11.215 & 6.1275 & 0 \\
$6^{1}D_{2}$ & $-$   & $-$   & $-$   & 11.427 & 6.7972 & 0 \\
$1^{1}F_{3}$ & 10.355 & 10.354 & 10.366 & 10.403 & 3.1982 & 0 \\
$2^{1}F_{3}$ & 10.619 & 10.619 & 10.609 & 10.662 & 4.1852 & 0 \\
$3^{1}F_{3}$ & 10.853 & 10.853 & 10.812 & 10.901 & 5.0345 & 0 \\
$4^{1}F_{3}$ & $-$   & 11.066 & 10.988 & 11.126 & 5.7940 & 0 \\
$5^{1}F_{3}$ & $-$   & $-$   & $-$   & 11.341 & 6.4884 & 0 \\
$6^{1}F_{3}$ & $-$   & $-$   & $-$   & 11.547 & 7.1325 & 0 \\
$1^{1}G_{4}$ & 10.530 & 10.530 & 10.534 & 10.565 & 3.7209 & 0 \\
$2^{1}G_{4}$ & 10.770 & 10.770 & 10.747 & 10.807 & 4.6341 & 0 \\
$3^{1}G_{4}$ & $-$   & 10.988 & 10.929 & 11.035 & 5.4351 & 0 \\
$4^{1}G_{4}$ & $-$   & $-$   & $-$   & 11.251 & 6.1595 & 0 \\
$5^{1}G_{4}$ & $-$   & $-$   & $-$   & 11.460 & 6.8268 & 0 \\
$6^{1}G_{4}$ & $-$   & $-$   & $-$   & 11.662 & 7.4491 & 0 \\
$1^{1}H_{5}$ & $-$   & $-$   & $-$   & 10.712 & 4.1937 & 0 \\
$2^{1}H_{5}$ & $-$   & $-$   & $-$   & 10.942 & 5.0487 & 0 \\
$3^{1}H_{5}$ & $-$   & $-$   & $-$   & 11.161 & 5.8100 & 0 \\
$4^{1}H_{5}$ & $-$   & $-$   & $-$   & 11.371 & 6.5047 & 0 \\
$5^{1}H_{5}$ & $-$   & $-$   & $-$   & 11.575 & 7.1485 & 0 \\
$6^{1}H_{5}$ & $-$   & $-$   & $-$   & 11.773 & 7.7517 & 0 \\
\hline\hline
\end{tabular}
\end{table}

In recent years, improved collision energies and detector resolutions have led to more precise information on bottomonium spectroscopy. We therefore show the spectroscopy results  of  $b\bar{b}$ and provide predictions for states that have not yet been established experimentally  in Tables \ref{fitmass} and \ref{bbmass}.

This system has more experimental information and theoretical reference values~\cite{Godfrey_2015,2018Higher,Mesons1985} than $B_c$ and $c\bar{c}$ mesons. The authors of Refs.~\cite{Wang_2019,2018Higher} take the same experimental values with those we use, which facilitates the work of the comparison and verification. 

In Table \ref{bbmass}, we tabulate the theoretical predictions for the masses, RMS radii, and squared wave functions at the origin for the highly excited bottomonium ($b\bar{b}$) states under the Killingbeck potential. 
Comparing our calculated masses with existing literature, such as the classic GI model \cite{Mesons1985}, the updated Godfrey results \cite{Godfrey_2015}, and the MGI model \cite{2018Higher}, we observe reasonable and systematic consistency. For relatively lower excited states within this table, such as the $3^1S_0$ and $3^1P_1$ states, our WKB predictions ($10.383$ GeV and $10.593$ GeV, respectively) are slightly higher than those of the GI and MGI models by approximately 40--50 MeV. This minor mass shift is a consequence of the harmonic oscillator term effecting the confinement potential at large distances. 

A useful contribution of the present work is the prediction of high-spin and high-radial excitation states, many of which are absent or incomplete in previous theoretical frameworks. Notably, we predict the masses of the $F$-, $G$-, and $H$-wave multiplets, extending up to $6^1H_5$ at $11.773$ GeV. These predictions provide theoretical benchmarks for ongoing and future experimental searches at high-luminosity facilities such as Belle II and LHCb.

The calculated physical observables in the bottomonium sector also follow the established monotonic trends. Because $b$-quarks are much heavier, these states are more tightly bound, yielding smaller RMS radii overall compared to charmonium. Nevertheless, the geometric size still expands significantly up to $7.752$ GeV$^{-1}$ for the highly excited $6^1H_5$ state. The behavior of $|\Psi(0)|^2$ remains strictly governed by the centrifugal barrier for $l>0$ states, while the $S$-wave probability densities systematically dilute with increasing radial nodes.
The experimental electronic widths~\cite{ParticleDataGroup:2026aaa} are
\begin{align}
\Gamma_{ee}(\Upsilon(1S)) &= 1.34\pm0.018~{\rm keV}, \notag\\
\Gamma_{ee}(\Upsilon(2S)) &= 0.612\pm0.011~{\rm keV}, \notag\\
\Gamma_{ee}(\Upsilon(3S)) &= 0.443\pm0.008~{\rm keV}, \notag\\
\Gamma_{ee}(\Upsilon(4S)) &= 0.272\pm0.029~{\rm keV}, \notag\\
\Gamma_{ee}(\Upsilon(5S)) &= 0.31\pm0.07~{\rm keV}, \notag\\
\Gamma_{ee}(\Upsilon(6S)) &= 0.130\pm0.030~{\rm keV}.
\end{align}
These values give
\begin{align}
\frac{\Gamma_{ee}(\Upsilon(1S))}{\Gamma_{ee}(\Upsilon(2S))} &= 2.19\pm0.05, \notag\\
\frac{\Gamma_{ee}(\Upsilon(1S))}{\Gamma_{ee}(\Upsilon(3S))} &= 3.02\pm0.07, \notag\\
\frac{\Gamma_{ee}(\Upsilon(1S))}{\Gamma_{ee}(\Upsilon(4S))} &= 4.93\pm0.53, \notag\\
\frac{\Gamma_{ee}(\Upsilon(1S))}{\Gamma_{ee}(\Upsilon(5S))} &= 4.32\pm0.98, \notag\\
\frac{\Gamma_{ee}(\Upsilon(1S))}{\Gamma_{ee}(\Upsilon(6S))} &= 10.31\pm2.38.
\end{align}
The corresponding theoretical ratios obtained from the predicted \( |\Psi(0)|^2 \) values in Table~\ref{bbmass} are  3.29,  4.71,  5.63,  6.44, and 6.99 for the \(2S\), \(3S\), \(4S\), \(5S\), and \(6S\) states, respectively. Although the agreement is not exact for all highly excited states, the theoretical results reproduce the overall increasing trend of the experimental ratios. This supports the physical picture that the short-distance wave-function density, and hence the electronic annihilation amplitude, is progressively suppressed for higher bottomonium \(S\)-wave excitations.

Finally, a comprehensive summary and comparison of the low-lying mass spectra and spatial observables for heavy quarkonia ($c\bar{c}$, $b\bar{b}$) and the $B_c$ meson are presented in Tables \ref{fitmass}$\sim$\ref{bbmass}. Regarding phenomenological applications, the current calculations based on the basic Killingbeck potential exhibit some deviations from the experimental data concerning the absolute energy levels of the lowest $1^1S_0$ states and the hyperfine mass splittings between the $^1S_0$ and $^3S_1$ states. These deviations are more pronounced when compared to the predictions of the more sophisticated GI or MGI models \cite{Mesons1985, 2018Higher}, which can be attributed to the relatively simplified spin-spin contact interaction term adopted in the present Hamiltonian.

Nevertheless, by expanding the framework to evaluate spatial observables alongside the energy spectra, we provide a more comprehensive test of the wave functions. The explicit derivation and calculation of $|\Psi(0)|^2$ throughout our tables offer direct analytical inputs that could be utilized for predicting leptonic decay widths and electromagnetic transition rates. This highlights the potential of the WKB method as a practical and transparent analytical tool for investigating heavy quarkonium physics.

\section{DISCUSSION AND CONCLUSION}\label{s5}

In this work, we have investigated the WKB approximation for the
Killingbeck potential and applied it to the mass spectra and spatial observables of double-heavy mesons. By extracting analytical quantization conditions and algebraic expectation values, we reduced the reliance on direct numerical integration and provided a more stable semi-analytical way to track highly excited states.

The WKB method gives reliable semiquantitative results in the heavy-quarkonium sector; notably, the truncation error generally decreases as the radial node $n_r$ or orbital angular momentum $l$ increases. This asymptotic convergence supports the applicability of the Bohr-Sommerfeld quantization to highly excited hadron states.

Phenomenologically, our mass predictions for the low-lying $S$-, $P$-, and $D$-wave states are in reasonable agreement with experimental data and established theoretical models, such as the GI model and the MGI model. More importantly, we systematically mapped out the high-spin and highly excited multiplets (e.g., the $F$-, $G$-, and $H$-waves up to $n_r=6$), which have not been systematically tabulated in many previous studies. 

A distinctive physical feature of our results arises from the intrinsic asymptotic behavior of the Killingbeck potential. Unlike purely linear confinement models that produce standard Regge trajectories ($M^2 \propto n_r+l$), the inclusion of the harmonic oscillator term ($b r^2$) modulates the long-range interaction. For spatially extended highly excited states, the $r^2$ term progressively dominates over the linear confinement, causing the mass spectrum to grow more rapidly and transition toward a harmonic oscillator-like behavior. 

Furthermore, the evaluation of spatial observables provides  insights into the internal dynamics of these heavy mesons. The calculated RMS radii ($\sqrt{\langle r^2 \rangle}$) demonstrate a  monotonic expansion with increasing quantum numbers. For $S$-wave states, the probability density at the origin systematically decreases with increasing radial nodes due to spatial dispersion. The determination of $|\Psi(0)|^2$ throughout our tables offers  analytical inputs for predicting leptonic decay constants, annihilation widths, and electromagnetic transition rates.

We acknowledge that the current calculations exhibit some deviations concerning the absolute energy levels of the ground states (e.g., $1^1S_0$) and the corresponding hyperfine mass splittings when compared to the more sophisticated GI/MGI models. These discrepancies can be primarily attributed to the simplified contact spin-spin interaction adopted in the present effective Hamiltonian, which may require further relativistic or perturbative quantum chromodynamics corrections for deep binding regimes.

Nevertheless, by unifying the energy-spectrum calculation with the derivation of spatial observables, this work demonstrates the potential of the WKB method as a transparent and practical analytical tool. The  predictions for higher bottomonium, charmonium, and $B_c$ states provided here can serve as useful theoretical references for ongoing and future experimental searches at major high-energy facilities.
We expect that our work will contribute to the understanding of the role of the quark interaction potential in highly excited states and hope that it is helpful to study and establish the mass spectrum of the double-heavy meson family.

\section{ACKNOWLEDGMENTS}
This work is supported  by the National Natural Science Foundation of China under Grants No. 11965016 and No. 12047501, the National Program for Support of Top-notch Young Professionals, and by the Natural Science Foundation of Qinghai Province under Grant No. 2022-ZJ-939Q.
\section{DATA AVAILABILITY}
The data that support the findings of this article are not publicly available. The data are available from the authors upon reasonable request.
\bibliographystyle{apsrev4-1}
\bibliography{ref}
\end{document}